\documentclass[11pt]{amsart}
\usepackage{amsfonts,latexsym,amssymb,dsfont,amsmath,etex,slashed,cancel}
\usepackage{amsthm}
\usepackage[dvipsnames]{pstricks}
\usepackage{pst-solides3d}
\usepackage{a4wide}
\usepackage{hyperref}

\hypersetup{
    colorlinks,
    linkcolor={red!50!black},
    citecolor={blue!50!black},
    urlcolor={blue!80!black}
}

\newcommand{\dc}{\overline}

\def\D{\slashed\nabla}

\newcommand{\K}{\mathcal{K}}
\newcommand{\nfs}{\,\cancel{\phantom{N}}\!\!\!\!\!\!\!\nabla\!\! f}
\newcommand{\DR}{\D_{\!\! R}}
\newcommand{\DL}{\D_{\!\! L}}
\newcommand{\DS}{\D_{\!\Sigma}}
\newcommand{\Sol}{\mathrm{Sol}}
\newcommand{\SOL}{\mathbf{Sol}}
\newcommand{\id}{{\ensuremath{\mathds{1}}}}
\newcommand{\dV}{\mathrm{dV}}
\newcommand{\dA}{\mathrm{dA}}
\newcommand{\Adach}{{\widehat{\mathrm{A}}}}
\newcommand{\ch}{\mathrm{ch}}
\newcommand{\Tr}{\mathrm{Tr}}
\newcommand{\nE}{\nabla^E}
\newcommand{\Ul}{\mathrm{U}(1)}
\newcommand{\He}{\mathrm{He}}

\newcommand{\<}{\langle}
\renewcommand{\>}{\rangle}

 \newtheorem{theorem}{Theorem}[section]

 \theoremstyle{definition}
 
 \newtheorem{exa}[theorem]{Example}

\DeclareMathOperator{\tr}{tr}

\allowdisplaybreaks
\title[A geometric formula for the chiral anomaly]{A rigorous geometric derivation of the chiral anomaly in curved backgrounds}

\author[C. B\"ar]{Christian B\"ar}
\address{Institute for Mathematics, Potsdam University, Karl-Liebknecht-Str.~24-25, 14476 Potsdam, Germany}
\email{baer@math.uni-potsdam.de}

\author[A. Strohmaier]{Alexander Strohmaier}
\address{Department of Mathematical Sciences,  Loughborough University,  Loughborough, Leicestershire, LE11 3TU, UK} 
\email{a.strohmaier@lboro.ac.uk}

\date{\today}

\begin{document}

\begin{abstract}
 We discuss the chiral anomaly for a Weyl field in a curved background and show that a novel index theorem
 for the Lorentzian Dirac operator can be applied to describe the gravitational chiral anomaly.
 A formula for the total charge generated by the gravitational and gauge field background is derived directly in Lorentzian signature 
  and in a mathematically rigorous manner. It contains a term identical to the integrand in the Atiyah-Singer index theorem and another term involving the $\eta$-invariant of the Cauchy hypersurfaces.
\end{abstract}
\maketitle

%%%%%%%%%%%%%%%%%%%%%%%%%%%%%%%%%%%%%%%%%%%%%%%%%%%%%%%%%%%%

\section{Introduction}

Anomalies in quantum field theory appear as a violation of conservation laws of currents, in other words, as currents that are classically preserved but whose
quantum counterparts are not. These anomalies are of direct physical significance. A prominent example is the chiral anomaly
(also ABJ-anomaly or axial anomaly). It explains the rate of decay of the neutral pion into two photons (\cite{Adler:1969aa,bell1969pcac}), 
$\pi_0 \to \gamma \gamma$. We would like to refer to the monograph \cite{bertlmann2000anomalies} for further details and applications in quantum field theory.
Apart from high energy physics, it has also been proposed that this anomaly can be observed more directly in crystals (see \cite{nielsen1983adler}, and also \cite{xiong2015signature} 
for recent experimental evidence). \\

\emph{Some historical remarks.}
Whereas anomalies were first discovered in perturbative computations in quantum field theory,
their appearance is related to the Atiyah-Singer index theorem. 
Indeed, the perturbative computation yields a term that looks precisely like the local Chern character form of the connection induced by an external electromagnetic field and perturbative computations on curved space-times
yield a term that is formally identical to the Atiyah-Singer integrand (\cite{kimura1969divergence,delbourgo1972gravitational}). 
However, these computations were performed in Lorentzian signature whereas the  Atiyah-Singer index theorem for Dirac operators is a theorem in Euclidean signature.

A direct computation of the divergence of the anomalous chiral current and the relation to the local index theorem was studied in Euclidean signature by Nielsen, R\"omer, and Schroer in \cite{nielsen-roemer-schroer1978}, and by Dowker  in \cite{MR0496159} in great detail. 
Another popular method using the Euclidean formulation of quantum field theory and path integrals is the Fujakawa method (\cite{MR758029}), which has also been applied in curved backgrounds (\cite{MR867029}, see also \cite{alvarez1985structure}). 
Both explanations of the relation to index theory can be seen only as formal manipulations because the process of Wick rotation, passing from Lorentzian to Euclidean signature, cannot be made sense of in general.

Another, more mathematically rigorous, way to understand anomalies in quantum field theory in external fields is via a careful analysis of the second quantization procedure
(\cite{MR0260291}). This was done
by Klaus and Scharf for fermions in an external field (\cite{MR0489514,MR0484140}). The one-particle time-evolution in an external field
can mix the negative and positive  energy solutions of the Dirac equation. This mixing can occur in such a way that the vacuum is mapped to a charged sector 
under the second quantized time-evolution (\cite{MR0489514,MR0484140,lundberg1976quasi,carey1982automorphisms}). 
The charge generated by this process relates to the Fredholm index of a certain
operator constructed out of the positive energy projection and the time-evolution operator. This is the point of view we will take in this paper. 

The fact that fermions and chiral charge can be created by a gravitational field was discovered by Gibbons in \cite{gibbons1979} (see also \cite{gibbons1980gravitational}). 
He computed this number in several models and showed that a Wick rotated Euclidean version was described by the Atiyah-Patodi-Singer index formula \cite{gibbons1980spectral}, i.e.\ by the Atiyah-Singer integral and a correction term involving the $\eta$-invariant of the boundary.
As pointed out in \cite{gibbons1979}, passing from the Euclidean formulation to the Lorentzian is highly problematic as only few Lorentzian manifolds admit an analytic continuation to a real Riemannian space which is non-singular. 
This is also mentioned by Lohiyah in \cite{lohiya1983anomalous} where the loss of chiral charge from the Reissner-Nordstr\"om black hole was computed directly but found to be consistent with the predictions of the Atiyah-Singer formula.
 \\

\emph{The present article.}
We will show in this paper that the gravitational chiral charge creation is given by the index of the Lorentzian Dirac operator with certain boundary conditions.
We use a novel index theorem for hyperbolic operators proved by the authors in \cite{Baer:2015aa} to show that this index is indeed, in analogy to the Atiyah-Patodi-Singer index theorem in Euclidean signature, given as a sum of the Atiyah-Singer integrand and a correction term involving the $\eta$-invariant.
This therefore establishes the formula directly for Lorentzian space-times and avoids the use of a Wick rotation.
This way we overcome the above mentioned problems with analytic continuation and provide a direct explanation why the Atiyah-Patodi-Singer formula yields  the correct result in the exactly solvable models.
Note that hyperbolic operators such as the Lorentzian Dirac operators have significantly different analytic properties from elliptic Dirac operators in Euclidean signature.
For example, the fact that the hyperbolic Dirac operator with spectral boundary conditions is a Fredholm operator and has a well-defined index at all, is due to the global propagation of singularities theorem and not to local regularity theory as in the elliptic case.

In order to make a precise mathematical statement about the created physical charge we work in the framework
of algebraic quantum field theory in curved space-times. 
This is natural as the notion of particle is not covariant and it is more appropriate in this context to speak about observables and states. 
We start by explaining in detail how the Weyl field is quantized in a globally hyperbolic space-time with spin structure. 
As is usual in quantum field theory in curved space-times we split this procedure into two parts (see for example \cite{wald1994quantum} for an introduction). 
The field algebra can be constructed in a functorial (covariant) manner from the field equation. 
This was first done by Dimock (\cite{MR637032}) for the Dirac field. 

In the second step we look at physical representations of the field algebra induced by states. 
The Hadamard condition singles out a class of states whose behavior resembles that of finite energy states in Minkowski space-time. 
In our context the important fact is that Hadamard states differ only by smooth integral kernels. 
Quantities that are a priory singular and need regularization, such as the expectation value of the energy momentum tensor or the electric current,  make sense as relative quantities between two Hadamard states. 
It is therefore possible to define the relative energy-momentum tensor and the relative current between two Hadamard states without the need for regularization.

A particular example of a Hadamard state can be defined if the space-time is ultrastatic in a neighborhood of a smooth spacelike Cauchy hypersurface. 
Near this Cauchy hypersurface a one parameter group of time-translations can be defined and it makes sense to define the vacuum state with respect to this time-evolution by employing the usual frequency splitting procedure.
We will think of this state as the vacuum state seen by an observer moving along the distinguished timelike Killing vector field near the Cauchy hypersurface.
If there are two such Cauchy hypersurfaces we can compare the two vacua. Due to the interaction with the gravitational background and the external gauge field between the two Cauchy hypersurfaces it can happen that the difference of the expectation values of the electric current between the two states is non-zero. 
The difference of the expectation values of the total charge operator turns out to be given essentially by the index of the Lorentzian Dirac operator with Atiyah-Patodi-Singer boundary conditions. 
We then apply the index theorem proved by the authors in \cite{Baer:2015aa} to obtain an explicit formula for the charge generated by the background fields.\\

\emph{More recent related work.}
An index formula for the scattering operator of the Dirac equation coupled to an external field was established by Matsui in \cite{matsui1987index,matsui1990index}.
Bunke and Hirschmann \cite{bunke1992index} generalized this to also allow time-dependent metric connections. 
In these results there is no $\eta$-correction term because they are related to the special case of our result where the future and past spacelike Cauchy hypersurfaces coincide so that the $\eta$-contributions cancel.

The computation of the divergence of the anomalous regularized current can also be carried out directly in Lorentzian signature using the Hadamard regularization.
This method makes use of the unique singularity structure of Hadamard states and subtracts the singular part from the singular current to make it a well-defined covariant and local object.
The computation of the divergence of the resulting current was done in \cite{zahn2015locally} for the mixed and pure gravitational anomalies and in \cite{MR2588824} for the trace anomaly that also can be described by an index theorem (\cite{christensen1979new}). 
The algebraic expressions coincide with the ones obtained in the Euclidean framework and those obtained in perturbative computations. The divergence of the anomalous current 
captures aspects of renormalization in curved space-times. However, the actual generation of charge in external fields is described by the index of the Dirac operator and the $\eta$-correction
term is vital to guarantee that the total charge generated is an integer.\\

\emph{Acknowledgments.}
It is our pleasure to thank Alan Carey, Klaus Fredenhagen, Harald Grosse, Stefan Hollands, Ken Richardson, and Christoph Stephan for very interesting and helpful discussion.
We would also like to thank Gary Gibbons for pointing out relevant references and for explaining to us the history of the chiral anomaly in the cosmological context.
Moreover, we are grateful to the anonymous referees for useful comments and for pointing out additional literature.
We thank \emph{Sonderforschungsbereich 647} funded by the \emph{Deutsche Forschungsgemeinschaft} for financial support.
The work for this paper was carried out during the program \emph{Modern Theory of Wave Equations} at the \emph{Erwin Schr\"odinger Institute} and we are grateful to the ESI for support and the hospitality during our stay.

\section{The setup}

We start by describing the classical setup which gives rise to the one-particle evolution.
We will consider massless fermions with internal symmetries in an external field on a curved space-time which satisfy a Dirac equation.
Let $(X,g)$ be an even-dimensional Lorentzian manifold where $g$ has signature $(+,-,\ldots,-)$.
We assume that $(X,g)$ is globally hyperbolic meaning that there exist Cauchy hypersurfaces so that there will be a well-posed initial value problem for the Dirac equation.
Then $(X,g)$ is time-orientable and we fix one time-orientation.
Moreover, we assume that $(X,g)$ is spatially compact, i.e.\ that the Cauchy hypersurfaces are compact.
We assume a spin structure on $X$ is given so that the complex spinor bundle $SX\to X$ is defined.

Finally, to model the internal symmetries, a Hermitian vector bundle $E\to X$ is fixed and, to describe an external gauge field, a compatible connection $\nE$ on $E$ is given.
Compatibility means that the Leibniz rule holds for $\nE$ and the scalar product on $E$.
For example, $E$ could be a Hermitian line bundle;
then $\nE$ is a $\Ul$-connection and can describe the electro-magnetic potential.

\subsection{The Dirac operator}
Spinors are sections of the bundle $SX\otimes E$.
Let $\nabla$ be the connection on $SX\otimes E$ induced by the Levi-Civita connection $\nabla^S$ on $SX$ and the connection $\nE$ on $E$, i.e.\ $\nabla_X(\sigma\otimes e)= (\nabla^S_X \sigma) \otimes e + \sigma\otimes \nE_Xe$.
The Dirac operator acts on spinors and is locally given by 
$$
i\D = ig^{\alpha\beta}\gamma_\alpha\nabla_\beta 
$$
where we used Einstein's summation convention and the coefficient matrices satisfy $\{\gamma_\alpha,\gamma_\beta\}=2g_{\alpha\beta}$.
Here and henceforth $\{\cdot,\cdot\}$ denotes the anticommutator.

Since the dimension of $X$ is even the spinor bundle splits into the subbundles of left-handed and right-handed spinors, $SX=S_LX\oplus S_RX$.
The Dirac operator interchanges chirality, i.e.\ with respect to the splitting $SX\otimes E=S_LX\otimes E\oplus S_RX\otimes E$ we have
$$
\D =
\begin{pmatrix}
0 & \DR \\
\DL & 0
\end{pmatrix}
.
$$
Note that there is no mass term on the diagonal.

The bundle $SX$ comes equipped with a natural nondegenerate but indefinite inner product $\<\cdot,\cdot\>$.
We use the convention that $\<\cdot,\cdot\>$ is antilinear in the first argument and linear in the second.
The subbundles $S_{L/R} X$ are isotropic with respect to this inner product.
The same is of course true for the induced inner products on $SX\otimes E$ and $S_{L/R} X\otimes E$.
Recall that the inner product on $E$ is positive definite.

The Dirac operator is formally selfadjoint with respect to $\<\cdot,\cdot\>$, i.e.\
\begin{equation}
\int_X \<i\D u,v\>\,\dV = \int_X \<u,i\D v\> \,\dV
\label{eq:Dformsa}
\end{equation}
where $u,v\in C^\infty_0(X;SX\otimes E)$ and $\dV$ is the volume element on $X$ induced by the Lorentzian metric.

The formally dual operator of $\DR:C^\infty(X;S_RX\otimes E)\to C^\infty(X;S_LX\otimes E)$ is the operator $\DR^*:C^\infty(X;S_L^*X\otimes E^*)\to C^\infty(X;S_R^*X\otimes E^*)$ on the dual bundle characterized by
\begin{equation}
\int_X (\DR^*v)(u)\dV = \int_X v(\DR u)\dV
\label{eq:Ddual}
\end{equation}
where $u\in C^\infty_0(X;S_RX\otimes E)$ and $v\in C^\infty_0(X;S_L^*X\otimes E^*)$.

\begin{exa}
If $E$ is a Hermitian line bundle, then the curvature of the dual connection on $E^*$ is the negative of that of $E$.
Up to a factor $i$, the curvature is a real $2$-form which describes the electro-magnetic field.
Locally, if we write $\nE_\alpha=\partial_\alpha+iA_\alpha$ then $\nabla^{E^*}_\alpha=\partial_\alpha-iA_\alpha$.
For the Dirac operators this means $i\D=i\gamma^\alpha(\partial_\alpha+iA_\alpha)=i\gamma^\alpha\partial_\alpha-A_\alpha\gamma^\alpha$ while $i\D^*=i\gamma^\alpha\partial_\alpha+A_\alpha\gamma^\alpha$.
Thus $i\D^*$ should be thought of as the operator with charge opposite to that of $i\D$.
\end{exa}

\subsection{Dirac conjugation}
For $u\in S_RX_x\otimes E_x$ define $\dc u\in S_L^*X_x\otimes E_x^*$ by $\dc u(v) = \<u,v\>$.
The map $u\mapsto\dc u$, $S_RX\otimes E \to S_L^*X\otimes E^*$ is antilinear.
Similarly we get an antilinear map $\dc\cdot:S_LX\otimes E \to S_R^*X\otimes E^*$ and the inverse maps $S_{R/L}^*X\otimes E^* \to S_{L/R} X\otimes E$ are also denoted by~$\dc\cdot$.
Equations~\eqref{eq:Dformsa} and \eqref{eq:Ddual} imply 
\begin{equation}
i\D^*\dc u=\dc{(i\D) u}. 
\label{eq:DTwiddel}
\end{equation}
Dirac conjugation is defined as the map 
\begin{equation}
\Gamma:S_L^*X\otimes E^* \oplus S_RX\otimes E \to S_L^*X\otimes E^* \oplus S_RX\otimes E,
\quad
u\oplus v \mapsto \dc v\oplus\dc u.
\label{eq:GammaDef}
\end{equation}
Clearly $\Gamma$ is antilinear and an involution, $\Gamma^2=\id$.

\subsection{The Cauchy problem}
Denote the space of smooth solutions of $\DR u=0$ by $\Sol(\DR):=\{u\in C^\infty(X;S_RX\otimes E) \mid \DR u=0\}$ and similarly for $\DL$, $\DR^*$, and $\DL^*$.
The Cauchy problem for the Dirac equation on globally hyperbolic manifolds is well posed (see e.g. \cite[Thm.~2.3]{MR637032}).
This means that for any smooth spacelike Cauchy hypersurface $\Sigma\subset X$ the restriction map
$$
\rho_\Sigma : \Sol(\DR) \to C^\infty(\Sigma;S_RX\otimes E),\quad
u \mapsto u|_\Sigma,
$$
is an isomorphism of topological vector spaces.
If $\Sigma$ and $\Sigma'$ are two smooth spacelike Cauchy hypersurfaces of $X$, then we put 
$$
U_{\Sigma',\Sigma}:=\rho_{\Sigma'}\circ(\rho_\Sigma)^{-1} : 
C^\infty(\Sigma;S_RX\otimes E)\to C^\infty(\Sigma';S_RX\otimes E).
$$
This evolution map $U_{\Sigma',\Sigma}$ extends to a unitary isomorphism 
\begin{equation}
U_{\Sigma',\Sigma}:L^2(\Sigma;S_RX\otimes E)\to L^2(\Sigma';S_RX\otimes E) \, .
\label{eq:U}
\end{equation}
Hence there is a unique Hilbert space completion $\SOL(\DR)$ of $\Sol(\DR)$ such that for each smooth spacelike Cauchy hypersurface $\Sigma\subset X$ the restriction map extends to a Hilbert space isometry
$$
\rho_\Sigma: \SOL(\DR)\to L^2(\Sigma;S_RX\otimes E),
$$
see \cite[Lemma~3.17]{MR3289848}.
We denote the scalar product on $\SOL(\DR)$ by $(\cdot,\cdot)$.
Similarly, we obtain the Hilbert space completions $\SOL(\DL)$, $\SOL(\DR^*)$, and $\SOL(\DL^*)$ of $\Sol(\DL)$, $\Sol(\DR^*)$, and $\Sol(\DL^*)$, respectively.

\subsection{The fermionic propagator}
Let $G_R: C^\infty_0(X;S_LX\otimes E) \to  C^\infty(X;S_RX\otimes E)$ be the difference between retarded and advanced fundamental solutions of $i\DR$.
Then $G_R$ maps onto the space of solutions $\mathrm{Sol}(\DR)$ of the Dirac equation $\DR u =0$.
This operator is sometimes called the fermionic propagator of $i\DR$.
Similarly, we obtain linear maps 
\begin{align*}
G_L : C^\infty_0(X;S_RX\otimes E) &\twoheadrightarrow  \Sol(\DL) \subset C^\infty(X;S_LX\otimes E), \\
G_{R,*} : C^\infty_0(X;S_R^*X\otimes E^*) &\twoheadrightarrow  \Sol(\DR^*) \subset C^\infty(X;S_L^*X\otimes E^*), \\
G_{L,*} : C^\infty_0(X;S_L^*X\otimes E^*) &\twoheadrightarrow  \Sol(\DL^*) \subset C^\infty(X;S_R^*X\otimes E^*).
\end{align*}
Equation \eqref{eq:DTwiddel} implies 
\begin{equation}
G_{R,*}\dc u = \dc{ G_Ru} \quad\mbox{ and }\quad G_{L,*}\dc v = \dc{ G_Lv}
\label{eq:GTwiddel}
\end{equation}
for all $u\in C^\infty_0(X;S_LX\otimes E)$ and $v\in C^\infty_0(X;S_RX\otimes E)$.
Using an integration by parts, we can express the scalar product on the Hilbert space $\SOL(\DR)$ as follows (compare \cite[Prop.~2.2]{MR637032})
\begin{equation}
(G_Rv,f) = -\int_X \< v,f\>\, \dV
\label{eq:ScalarProduct}
\end{equation}
where $v\in C^\infty_0(X;S_LX\otimes E)$ and $f\in\SOL(\DR)$.
Analogous formulas hold for the scalar products on $\SOL(\DL)$, $\SOL(\DR^*)$, and $\SOL(\DL^*)$, respectively.

By \cite[Thm.~4.3]{MR3302643} the fermionic propagators extend to operators on distributional sections, e.g.\
$$
G_R: C^{-\infty}_0(X;S_LX\otimes E) \to  C^{-\infty}(X;S_RX\otimes E).
$$
The image is precisely the space of distributional solutions of the Dirac equation.

\subsection{Fermionic propagator and Cauchy problem}

Let $\Sigma\subset X$ be a spacelike smooth Cauchy hypersurface with future-directed unit normal field $n_\Sigma$.
Any $u\in C^{-\infty}(\Sigma; S_LX\otimes E)$ gives rise to a distribution $u\delta_\Sigma$ on $X$ via $(u\delta_\Sigma)(v)=\int_\Sigma (\rho_\Sigma v)(u)\,\dA$ for all $v\in C^\infty(X;S_L^*X\otimes E^*)$.
Since the support of $u\delta_\Sigma$ is contained in $\Sigma$ and hence compact, we can apply $G_R$ to $u\delta_\Sigma$.
An argument using integration by parts shows that $f=G_R(u\delta_\Sigma)\in C^{-\infty}(X;S_RX\otimes E)$ is the solution of the Cauchy problem 
\begin{equation}
i\DR f=0 \quad\mbox{ and }\quad \rho_\Sigma f=\slashed n_\Sigma u.
\label{eq:Cauchy}
\end{equation}
Here $\slashed n_\Sigma=n_\Sigma^\alpha\gamma_\alpha$ denotes Clifford multiplication by $n_\Sigma$ in accordance with Feynman's slash convention.
The restriction $G_R:L^2(\Sigma; S_LX\otimes E)\subset C^{-\infty}_0(X;S_LX\otimes E)\to\SOL(\DR)$ is therefore the inverse of the Hilbert space isometry $\slashed n_\Sigma\rho_\Sigma:\SOL(\DR)\to L^2(\Sigma; S_LX\otimes E)$ and hence is itself a Hilbert space isometry.

\section{States and the chiral anomaly}

In order to describe the quantized Weyl field we start with the construction of the field algebra.
It is essentially the CAR algebra associated with the space of solutions of the Dirac equation. 
More precisely,  let $\K$ be the Hilbert space sum $\mathbf{Sol}(\DR^*) \oplus \mathbf{Sol}(\DR)$.
We denote the scalar product by $(\cdot,\cdot)_\K$. 
This Hilbert space comes with the Dirac conjugation $\Gamma: \K \to \K$ as defined in \eqref{eq:GammaDef}.
Then the selfdual CAR algebra associated with the pair $(\K, \Gamma)$ is the unital $*$-algebra generated by symbols $B(f)$ where $f\in\K$ and relations
 \begin{gather}
  f \mapsto  B(f) \; \textrm{is complex linear},\label{eq:CAR1}\\
  \{B(f),B(g)\}= (\Gamma f, g )_\K,\label{eq:CAR2}\\
   B(f)^*=B(\Gamma f).\label{eq:CAR3}
 \end{gather}
The CAR algebra admits a unique $C^*$-norm and we will denote its $C^*$-completion by $\mathcal{A}$.
This is our field algebra.

\subsection{The field operators}
We define the field operators by
 \begin{align*}
 \Psi:C^\infty_0(X;S_R^*X\otimes E^*)\to \mathcal{A}, \quad &\Psi(u)= B(G_{R,*}u \oplus 0),\\
 \dc{\Psi}:C^\infty_0(X;S_LX\otimes E)\to \mathcal{A}, \quad &\dc{\Psi}(v)=B(0 \oplus G_Rv).
 \end{align*} 
Relations \eqref{eq:CAR1}--\eqref{eq:CAR3} together with \eqref{eq:GTwiddel} and \eqref{eq:ScalarProduct} imply that
 \begin{gather}
  u \mapsto  \Psi(u) \; \textrm{is complex linear},\notag\\
  v \mapsto  \dc{\Psi}(v) \; \textrm{is complex linear},\notag\\ 
   \{\Psi(u_1),\Psi(u_2)\}=0, \quad \{\dc{\Psi}(v_1),\dc{\Psi}(v_2)\}=0, \notag\\
   \{\dc{\Psi}(v),\Psi(u)\}=- i\int_X\<\dc v, G_{R,*}u\>\, \dV,\notag\\
   \Psi(u)^*=\dc{\Psi}(\dc u),\notag\\
   \DR \Psi=0, \quad \DR^*\dc{\Psi}=0. \label{eq:PsiDirac}
 \end{gather}
Here $\DR \Psi=0$ is to be understood in the distributional sense, i.e.\ $\Psi(\DR^*u)=0$ for all $u\in C^\infty_0(X;S_L^*X\otimes E^*)$ and similarly $\dc{\Psi}(\DR v)=0$ for all $v\in C^\infty_0(X;S_RX\otimes E)$.

\subsection{The \textit{n}-point functions}
A state $\omega$ on this algebra is of course uniquely determined by its $n$-point functions
$$
 \omega_n(f_1,\ldots,f_n) :=\omega(B(f_1) \cdots B(f_n)).
$$
This also defines a distributional section $\tilde \omega_n \in C^{-\infty}(X^n; \boxtimes^n (S_RX \otimes E \oplus S_L^*X \otimes E^*))$ on the $n$-fold Cartesian product of $X$ by
$$
 \tilde \omega_n(u_1 \otimes \cdots \otimes u_n):=\omega_n(\mathbf{G}u_1,\ldots,\mathbf{G}u_n)
$$
where we have used the notation $\mathbf{G} = G_{R,*} \oplus G_R$.
We will also refer to this distribution as the $n$-point function.

A state is called quasi-free if for all $n=1,2,3,\ldots$ and $f_i \in \K$
\begin{gather*}
 \omega_{2n-1}(f_1,\ldots,f_{2n-1}) =0,\\
 \omega_{2n}(f_1,\ldots,f_{2n}) = (-1)^{\frac{n(n-1)}{2}} \sum \mathrm{sgn}(s) \sum_{j=1}^n \omega_2(f_{s(j)}, f_{s(j+n)})
\end{gather*}
where the sum is taken over all permutations $s$ of $\{1,\ldots,2n\}$ such that
\begin{gather*}
 s(1) < s(2) < \cdots < s(n),\quad
 s(j) < s(j+n).
\end{gather*}
This means that the state is completely determined by its two-point function.
\subsection{Hadamard forms and relative currents}
A  two-point function is said to be of Hadamard form
if its wavefront set $\mathrm{WF}(\tilde \omega_2)$ satisfies
$$
 \mathrm{WF}(\tilde \omega_2) \subset \{ (x_1,\xi_1,x_2,\xi_2) \in T^*(X \times X) \mid (x_1,\xi_1) \sim (x_2,-\xi_2),\;\xi_2\; \textrm{is future directed}  \},
$$
where $(x_1,\xi_1) \sim (x_2,\xi_2)$ means that these vectors are in the same orbit of the geodesic flow.
It is known (see e.g. \cite{MR1860415,hollands2001hadamard} in the case of the Dirac field) 
 that Hadamard  forms are unique up to smooth kernels, i.e.\ if $\omega_1$ and $\omega_2$ are states with two-point functions of Hadamard form then$$
\tilde \omega_{1,2}
:=\tilde \omega_1 - \tilde \omega_2 
\in C^\infty(X \times X;  \boxtimes^2(S_R X \otimes E \oplus S_L^*X \otimes E^*)).
$$ 
Note that $ \tilde \omega_{1,2}$ is a smooth bi-solution of the Dirac equation. 
We can then define the \emph{relative current} $J^{\omega_1,\omega_2} \in \Omega^1(X)$ as follows. Given a state $\omega$ we define
$\hat S_\omega: C^\infty_0(X;S_R^* X \otimes E^*) \to C^{-\infty}(X;S_L^* X \otimes E^*)$ as the unique operator
such that
$$
\omega(\dc{\Psi}(v) \Psi(u))   = (\hat S_\omega u)(v)
$$
for all $v\in C^\infty_0(X;S_LX\otimes E)$.
Note that if $\omega$ has Hadamard form then we actually have
$\hat S_\omega: C^\infty_0(X;S_R^* X \otimes E^*) \to C^{\infty}(X;S_L^* X \otimes E^*)$.
We put $\hat S := \hat S_{\omega_1} - \hat S_{\omega_2}$. 
If both $\omega_1$ and $\omega_2$ have two-point functions of Hadamard form then the operator $\hat S$ has smooth integral kernel. 
We denote the integral kernel of $\hat S$ evaluated at $(x,y)\in M\times M$ by $\hat s(x,y)\in \mathrm{Hom}(S_R^*X_y\otimes E_y^*,S_L^*X_x\otimes E^*_x)$.
Moreover, by \eqref{eq:PsiDirac}, we obtain
\begin{equation}
\DR^*\circ \hat S = \hat S\circ \DR^*  =0.
\label{eq:SDirac}
\end{equation}
For $\xi\in TX_x$ denote Clifford multiplication with $\xi$ by $\slashed\xi\in \mathrm{Hom}(S_RX_x\otimes E_x,S_LX_x\otimes E_x)$ and its dual by $\slashed\xi^*\in \mathrm{Hom}(S_L^*X_x\otimes E_x^*,S_R^*X_x\otimes E_x^*)$.
Now $\slashed\xi^*\circ \hat s(x,x)\in \mathrm{End}(S_R^*X_x\otimes E_x^*)$ and we can set
$$
J^{\omega_1,\omega_2}(\xi) := \tr (\slashed\xi^*\circ \hat s(x,x)).
$$
Since $\hat s$ is smooth on $M\times M$, this defines a smooth one-form $J^{\omega_1,\omega_2} \in \Omega^1(X)$.
In physics terminology, one could write this definition as
$$
 J^{\omega_1,\omega_2}_\mu(x)
=
\lim_{y \to x} \left( \omega_1(\dc{\Psi}^{\,\dot A}(x) (\gamma_\mu)_{\dot A}^B \Psi_B(y)) - 
 \omega_2(\dc{\Psi}^{\,\dot A}(x) (\gamma_\mu)_{\dot A}^B \Psi_B(y))\right),
$$
where the Einstein summation convention was used on the dotted spinor index $\dot A$ and spinor index $B$.
The relative current can be thought of as the expectation value $\omega_2(:\!\!J_\mu(x)\!\!:)$ of the normally ordered current operator
$:\!\!J_\mu(x)\!\!:$, where the normal ordering has been done with respect to state $\omega_1$.
It follows directly from the definitions that
$$
[\slashed\nabla,f]=\nfs 
\quad\mbox{ and }\quad
[f,\slashed\nabla^*]=\nfs^{\,*} 
$$
where $f\in C^\infty(X)$ is a function and $\nabla\! f$ its gradient vector field.
For any smooth compactly supported function $f \in C^\infty_0(X)$ we get, using \eqref{eq:SDirac} twice,
$$
\int_X f\delta J^{\omega_1,\omega_2} \,\dV
=
\int_X J^{\omega_1,\omega_2}(\nabla\! f)\,\dV 
=
\Tr(\nfs^{\,*}\hat S)
=
\Tr (f\DR^* \hat S -  \DR^* f \hat S ) 
=
-\Tr(f \hat S\DR^*)
=
0.
$$
Here we have used the canonical trace $\Tr$ on the algebra of integral operators whose integral kernel have compact support in the first variable.
Its  properties are described in Appendix~\ref{appendix:trace}.

This shows $\delta J^{\omega_1,\omega_2}=0$.
If $\Sigma$ is a spacelike smooth Cauchy hypersurface and $n_\Sigma$ the future-directed unit normal vector field along $\Sigma$, then we can integrate the relative current $J^{\omega_1,\omega_2}(n_\Sigma)$ along the Cauchy hypersurface. 
Since the current is co-closed the integral does not depend on the choice of Cauchy hypersurface and defines the \emph{relative right-handed charge} $Q_R^{\omega_1,\omega_2}$ between two states
\begin{equation}
 Q_R^{\omega_1,\omega_2} := \int_\Sigma J^{\omega_1,\omega_2}(n_\Sigma(x)) \dA(x).
\label{eq:Q}
\end{equation}

\subsection{Fock representations}

If $\mathcal{H}$ is a Hilbert space the fermionic Fock space $\mathcal{F}(\mathcal{H})$ is defined to be the Hilbert space direct sum
$$
 \mathcal{F}(\mathcal{H})=\bigoplus_{k=0}^\infty {\Lambda}^{k} \mathcal{H}.
$$
As usual the vector $1 \in \mathbb{C} = \Lambda^0\mathcal{H} \subset \mathcal{F}(\mathcal{H})$ is denoted by $\Omega$ and is called the \emph{vacuum vector}.
On this space we have the usual creation and annihilation operators
\begin{gather*}
 a^\dagger(v) h = v \wedge h,\\
 a(v) h = \iota_v  h.
\end{gather*}
Note that $a^\dagger(v)$ is linear in $v$, whereas $a(v)$ is antilinear in $v$.

Let $P: \K \to \K$ be an orthogonal projection such that
$$
 P + \Gamma P \Gamma = \id.
$$
This induces an orthogonal splitting $\K = \mathcal{H} \oplus\Gamma {\mathcal{H}}$ where $\mathcal{H}= P \K$.
We obtain a faithful $*$-representation $\pi$ of $\mathcal{A}$ on the Fock space $\mathcal{F}(\mathcal{H})$ by setting
$$
 \pi(B(f)) := a^\dagger(P f) + a( P \Gamma f).
$$
This representation is called the Fock representation. 
One checks that for such a representation the state
$$
 \omega(\cdot )= \< \Omega, \cdot \;\Omega \>
$$
is a quasifree state and the Fock representation is canonically isomorphic to the GNS representation of $\omega$.
See \cite{MR0295702} or \cite{MR1312612} for a detailed discussion of CAR algebras and Fock representations.

\subsection{Fock representations constructed near $\mathbf{\Sigma}$}

If $\Sigma$ is a spacelike smooth Cauchy hypersurface then the restriction map 
$$
\rho_\Sigma: \mathcal{K}=\SOL(\DR^*)\oplus\SOL(\DR) \to 
\K_\Sigma
:=
L^2(\Sigma;S_L^*X \otimes E^*)\oplus L^2(\Sigma;S_RX \otimes E) 
$$ 
is an isomorphism of Hilbert spaces. 
We can therefore identify the Hilbert space $\K$ with conjugation $\Gamma$ with $\K_\Sigma$ and conjugation $\Gamma|_\Sigma$. 
Moreover, we can identify $S_RX|_\Sigma$ with $S\Sigma$, the spinor bundle of $\Sigma$.
Using the isomorphism $\slashed n_\Sigma:S_LX|_\Sigma\to S_RX|_\Sigma$ we also get an identification of $S_LX|_\Sigma$ with $S\Sigma$.
The Dirac operator $\DS$ on $\Sigma$ anticommutes with $\slashed n_\Sigma$ and is selfadjoint with respect to the positive definite scalar product $\<\slashed n_\Sigma \cdot,\cdot\>$ on $S\Sigma$.
This implies 
\begin{equation}
\DS\dc v=-\dc{\DS^* v}.
\label{eq:DSigma}
\end{equation}
In particular, conjugation $\dc\cdot$ maps $\DS$-eigenspaces for positive eigenvalues to $\DS^*$-eigenspaces for negative eigenvalues and vice versa.
We consider the spectral projectors
$$
p_\ge(\DS) := \chi_{[0,\infty)}(\DS)
\quad\mbox{ and }\quad
p_>(\DS) := \chi_{(0,\infty)}(\DS) \, .
$$
Similarly, we define $p_\le(\DS)$, $p_<(\DS)$, and the corresponding projectors for $\DS^*$.
Now 
$$
P:=p_\ge(\DS^*) \oplus p_>(\DS)
$$ 
is an orthogonal projection on $\K_\Sigma$ and \eqref{eq:DSigma} implies $P+\Gamma P\Gamma=\id$.
Its Fock representation will then be modeled on the one-particle Hilbert space
$$
\mathcal{H}_\Sigma = \mathcal{H}_\Sigma^+ \oplus   \mathcal{H}_\Sigma^-
$$
where
\begin{align*}
\mathcal{H}_\Sigma^+&= p_\ge(\DS^*) L^2(\Sigma;S^*\Sigma \otimes E^*), \\
\mathcal{H}_\Sigma^-&= p_>(\DS)L^2(\Sigma;S\Sigma \otimes E)=\dc{p_<(\DS^*) L^2(\Sigma;S^*\Sigma \otimes E^*)} \, .
\end{align*}
The space $\mathcal{H}_\Sigma^+$ is the one-particle space for particles and $\mathcal{H}_\Sigma^-$ is the one-particle space for antiparticles.
The Fock space then splits into a tensor product
$$
 \mathcal{F}(\mathcal{H}_\Sigma)= \mathcal{F}(\mathcal{H}_\Sigma^+) \hat \otimes \mathcal{F}(\mathcal{H}_\Sigma^-).
$$
As usual we denote by $c$ and $c^\dagger$ the creation and annihilation operators on $\mathcal{F}(\mathcal{H}_\Sigma^+)$ and by $d$ and $d^\dagger$ the creation and annihilation operators on $\mathcal{F}(\mathcal{H}_\Sigma^-) $. 
Each of them acts on the entire Fock space by extending it by the identity map on the other tensor factor.
The representation of $\mathcal{A}$ in this Fock representation is explicitly given by
\begin{gather*}
\pi_\Sigma(\Psi(u)) 
= 
c^\dagger\big(p_\ge(\DS^*) \rho_\Sigma G_{R,*} u\big) + d\big(p_>(\DS) \rho_\Sigma G_{R}\dc u\big),\\
\pi_\Sigma(\dc{\Psi}(v)) 
= 
c\big(p_\ge(\DS^*) \rho_\Sigma G_{R,*} \dc v\big) + d^\dagger\big(p_<(\DS)\rho_\Sigma G_{R} v\big).
\end{gather*}
The notation
$$
\pi_\Sigma(B(f)) = a^\dagger( P \rho_\Sigma f) + a(P \Gamma \rho_\Sigma f)
$$
with the substitution $f=\mathbf{G}(u\oplus v)$ is however more compact.

It is well known (see for example \cite[Sec.~4]{MR2273514}) that the vacuum expectation value $\omega_\Sigma$ with respect to this representation is a Hadamard state, i.e.\ its two-point function is of Hadamard form, if the metric on $X$ and $\nE$ are of product type near $\Sigma$. 
This means that a neighborhood of $\Sigma$ can be identified with $(-\varepsilon,\varepsilon)\times\Sigma$ in such a way that $\Sigma$ corresponds to $\{0\}\times\Sigma$, that the metric takes the form $dt^2-g_{\Sigma}$ where $g_{\Sigma}$ is independent of $t$ and that $\nE$ is the pull-back of its restriction to $\Sigma$ under the projection $(-\varepsilon,\varepsilon)\times\Sigma\to\Sigma$.

\begin{pspicture}(-7,-2.8)(6,2.4)
\psset{viewpoint=-30 10 15, Decran=30, lightsrc=-20 20 15}
\defFunction{Lorentz}(u,v)
 {u Cos 0.4 mul 0.6 add v Cos mul}
 {u Cos 0.4 mul 0.6 add v Sin mul}
 {u}
\defFunction{Lorentz2}(u,v)
 {u 1 add Cos -1 mul 2 add v Cos mul}
 {u 1 add Cos -1 mul 2 add v Sin mul}
 {u}
\defFunction{HalsDick}(u,v)
 {v Cos}
 {v Sin}
 {u}
\defFunction{Splus}(v)
 {v Cos}
 {v Sin}
 {-0.5}

\psSolid[object=surfaceparametree,
        base=-2 -1 0 2 pi mul,
        incolor=blue!70,
        fillcolor=RoyalBlue!70,
        opacity=0.7,
        function=Lorentz2,
        ngrid=120 180,
%        ngrid=12 18,
        grid=false]%
\psSolid[object=surfaceparametree,
        base=-1 0 0 2 pi mul,
        incolor=RoyalBlue!70,
        fillcolor=white,
        opacity=0.7,
        function=HalsDick,
        ngrid=20 180,
%        ngrid=2 18,
        grid=false]%
\psSolid[object=surfaceparametree,
        base=0 2 0 2 pi mul,
        incolor=blue!70,
        fillcolor=yellow,
        opacity=0.7,
        function=Lorentz,
        ngrid=20 180,
%        ngrid=4 36,
        grid=false]%
\psSolid[object=courbe,
        range=1.3 4.22,
        r=0,
        ngrid=360,
        linecolor=black,
        linewidth=0.02,
        function=Splus]%
       
\psPoint(-3,0.5,1){X}
\uput[u](X){\psframebox*[framearc=.3]{$X$}}
\psPoint(-3,-0.6,0.1){S}
\uput[u](S){\psframebox*[framearc=.3]{$\Sigma$}}
\end{pspicture}

\begin{center}
\textbf{Fig.~1.}
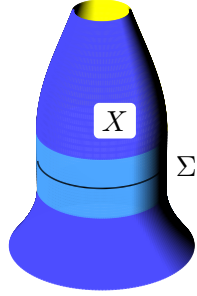
\emph{Manifold $X$ with product structure near $\Sigma$}
\end{center}

In this case this representation is thought of as the preferred vacuum representation of an observer on $\Sigma$.

\subsection{Integral kernels of spectral projectors}

As before let $\hat S_\Sigma$ be the unique operator $C^\infty_0(X;S_R^* X \otimes E^*) \to C^\infty(X;S_L^* X \otimes E^*)$ such that
$$
 \omega_\Sigma(\dc{\Psi}(v) \Psi(u))   = \int_X \<\dc v, \hat S_\Sigma u \>\,\dV
$$
and $\hat s_\Sigma$ its integral kernel.
Now let $u\in L^2(\Sigma;S_R^*X\otimes E^*)$.
Then $G_{R,*}(u\delta_\Sigma)\in\SOL(\DR^*)$ so that $\Psi(u)=B(G_{R,*}(u\delta_\Sigma)\oplus 0)$ is defined.
Similarly, $\dc\Psi(v)$ is defined for $v\in L^2(\Sigma; S_LX\otimes E)$.
We compute
\begin{align*}
\omega_\Sigma(\dc{\Psi}(v) \Psi(u))
&\stackrel{\phantom{\eqref{eq:Cauchy}}}{=}
\omega_\Sigma(B(0\oplus G_R(v\delta_\Sigma))B(G_{R,*}(u\delta_\Sigma)\oplus 0)) \\
&\stackrel{\phantom{\eqref{eq:Cauchy}}}{=}
\left(\Omega,\pi_\Sigma(B(0\oplus G_R(v\delta_\Sigma))B(G_{R,*}(u\delta_\Sigma)\oplus 0))\Omega\right)_{\mathcal{F}(\mathcal{H}_\Sigma)} \\
&\stackrel{\phantom{\eqref{eq:Cauchy}}}{=}
\left(\pi_\Sigma(B(\dc{G_R(v\delta_\Sigma)}\oplus 0))\Omega,\pi_\Sigma(B(G_{R,*}(u\delta_\Sigma)\oplus 0))\Omega\right)_{\mathcal{F}(\mathcal{H}_\Sigma)} \\
&\stackrel{\phantom{\eqref{eq:Cauchy}}}{=}
\left(P\rho_\Sigma(\dc{G_R(v\delta_\Sigma)}\oplus 0)\wedge\Omega,P\rho_\Sigma(G_{R,*}(u\delta_\Sigma)\oplus 0)\wedge\Omega\right)_{\mathcal{F}(\mathcal{H}_\Sigma)} \\
&\stackrel{\eqref{eq:Cauchy}}{=}
\left(p_\ge(\DS^*)(\dc{\slashed n_\Sigma v}),p_\ge(\DS^*)(\slashed n_\Sigma^* u)\right)_{\mathcal{H}_\Sigma^+} \\
&\stackrel{\phantom{\eqref{eq:Cauchy}}}{=}
\left(\slashed n_\Sigma^* \dc{v},p_\ge(\DS^*)(\slashed n_\Sigma^* u)\right)_{L^2(\Sigma)}  \\
&\stackrel{\phantom{\eqref{eq:Cauchy}}}{=}
\int_\Sigma \<\dc{v},p_\ge(\DS^*)(\slashed n_\Sigma^* u)\> \,\dA  \, .
\end{align*}
In other words, the integral kernel of the projector $p_\ge(\DS^*)$ coincides with $\hat s_\Sigma(y,x)\slashed n_\Sigma^*(x)$, restricted to $(y,x)\in\Sigma\times\Sigma$.

More generally, let $\Sigma_1$ and $\Sigma_2$ be two spacelike smooth Cauchy hypersurfaces and let $U_{\Sigma_1,\Sigma_2}$ the unitary evolution operator defined in \eqref{eq:U}.
For $u\in L^2(\Sigma_1;S_R^*X\otimes E^*)$ we have by \eqref{eq:Cauchy} that $G_{R,*}((\slashed n_{\Sigma_2}^*U_{\Sigma_1,\Sigma_2}^{-1}u)\delta_{\Sigma_2})=G_{R,*}(\slashed n_{\Sigma_1}^* u\delta_{\Sigma_1})$ and hence $\Psi(\slashed n_{\Sigma_2}^*U_{\Sigma_1,\Sigma_2}^{-1}\slashed n_{\Sigma_1}^*u)=\Psi(u)$ and similarly for $\dc\Psi$.
If $v\in L^2(\Sigma_1;S_LX\otimes E)$ we have on the one hand
$$
\omega_{\Sigma_2}(\dc\Psi(v)\Psi(u)) = \int_{\Sigma_1\times\Sigma_1}\<\dc v(y),\hat s_{\Sigma_2}(y,x)u(x)\> \dA(x)\dA(y)
$$
and on the other hand
\begin{align*}
\omega_{\Sigma_2}(\dc\Psi(v)\Psi(u))
&=
\omega_{\Sigma_2}(\dc\Psi(\slashed n_{\Sigma_2}U_{\Sigma_1,\Sigma_2}^{-1}\slashed n_{\Sigma_1}v)\Psi(\slashed n_{\Sigma_2}^*U_{\Sigma_1,\Sigma_2}^{-1}\slashed n_{\Sigma_1}^*u)) \\
&=
\left(\slashed n_{\Sigma_2}^* \dc{\slashed n_{\Sigma_2}U_{\Sigma_1,\Sigma_2}^{-1}\slashed n_{\Sigma_1}v},p_\ge(\D_{\Sigma_2}^*)(\slashed n_{\Sigma_2}^* \slashed n_{\Sigma_2}^*U_{\Sigma_1,\Sigma_2}^{-1}\slashed n_{\Sigma_1}^*u)\right)_{L^2(\Sigma_2)}\\
&=
\big(U_{\Sigma_1,\Sigma_2}^{-1}\slashed n_{\Sigma_1}^* \dc{v},p_\ge(\D_{\Sigma_2}^*)(U_{\Sigma_1,\Sigma_2}^{-1}\slashed n_{\Sigma_1}^*u)\big)_{L^2(\Sigma_2)}\\
&=
\big(\slashed n_{\Sigma_1}^*\dc{v},U_{\Sigma_1,\Sigma_2} p_\ge(\D_{\Sigma_2}^*)(U_{\Sigma_1,\Sigma_2}^{-1}\slashed n_{\Sigma_1}^* u)\big)_{L^2(\Sigma_1)}\\
&=
\int_{\Sigma_1}\big\<\dc{v},U_{\Sigma_1,\Sigma_2}p_\ge(\D_{\Sigma_2}^*)U_{\Sigma_1,\Sigma_2}^{-1}\slashed n_{\Sigma_1}^*u\big\> \, \dA \, .
\end{align*}
Hence the integral kernel of $U_{\Sigma_1,\Sigma_2}\circ p_\ge(\D_{\Sigma_2}^*)\circ U_{\Sigma_1,\Sigma_2}^{-1}$ coincides with $\hat s_{\Sigma_2}(y,x)\slashed n_{\Sigma_1}^*(x)$, restricted to $(y,x)\in\Sigma_1\times\Sigma_1$.

\subsection{Relative charge between Fock states associated to Cauchy hypersurfaces}

Now we consider the following situation:
let $\Sigma_1$ and $\Sigma_2$ be two spacelike smooth Cauchy hypersurfaces with $\Sigma_1$ lying in the past of $\Sigma_2$.
We assume that the metric of $X$ and the connection $\nE$ of $E$ have product structure near both hypersurfaces.
We compute the relative charge for the Fock states $\omega_{\Sigma_1}$ and $\omega_{\Sigma_2}$:
\begin{align*}
Q_R^{\omega_{\Sigma_1},\omega_{\Sigma_2}}
&=
\int_{\Sigma_1} J^{\omega_{\Sigma_2},\omega_{\Sigma_1}}(n_{\Sigma_1})\, \dA \\
&=
\int_{\Sigma_1} \tr\big(\slashed n_{\Sigma_1}^*(\hat s_{\Sigma_1}(x,x)-\hat s_{\Sigma_2}(x,x))\big) \,\dA(x) \\
&=
\int_{\Sigma_1} \tr\big((\hat s_{\Sigma_1}(x,x)-\hat s_{\Sigma_2}(x,x))\slashed n_{\Sigma_1}^*\big) \,\dA(x) \\
&=
\Tr\big(p_\ge(\D_{\Sigma_1}^*) - U_{\Sigma_1,\Sigma_2}p_\ge(\D_{\Sigma_2}^*)U_{\Sigma_1,\Sigma_2}^{-1}\big) \, .
\end{align*}
It was shown in \cite[Thm.~6.5]{Baer:2015aa} that the operator $p_\ge(\D_{\Sigma_1}^*) - U_{\Sigma_1,\Sigma_2}p_\ge(\D_{\Sigma_2}^*)U_{\Sigma_1,\Sigma_2}^{-1}$ has a smooth integral kernel. 
In particular, the operator is of trace class.
It now follows from \cite[Thm.~4.1]{MR1262254} that the trace is an integer and equals an index, namely putting $U=U_{\Sigma_1,\Sigma_2}$ and $p_j=p_\ge(\D_{\Sigma_j}^*)$,
\begin{align*}
Q_R^{\omega_{\Sigma_1},\omega_{\Sigma_2}}
&=
\mathrm{ind}\big[Up_2U^{-1}p_1:p_1L^2(\Sigma_1)\to Up_2U^{-1}L^2(\Sigma_1)\big] \\
&=
\mathrm{ind}\big[p_2U^{-1}p_1:p_1L^2(\Sigma_1)\to p_2U^{-1}L^2(\Sigma_1)\big] \, .
\end{align*}
By Theorem~4.1 and the concluding remark in \cite{Baer:2015aa} this index is given by 
\begin{align}
Q_R^{\omega_{\Sigma_1},\omega_{\Sigma_2}}
&= 
-\int_M \Adach\wedge\ch(\nE) 
+\frac{h(\D_{\Sigma_1}) + h(\D_{\Sigma_2})+\eta(\D_{\Sigma_1})-\eta(\D_{\Sigma_2})}{2} - h(\D_{\Sigma_2}) \label{eq:Q1}\\
&= 
-\int_M \Adach\wedge\ch(\nE) 
+\frac{h(\D_{\Sigma_1}) - h(\D_{\Sigma_2})+\eta(\D_{\Sigma_1})-\eta(\D_{\Sigma_2})}{2} \, .
\end{align}
Here $M$ is the region between $\Sigma_1$ and $\Sigma_2$, i.e.\ $M=J^+(\Sigma_1)\cap J^-(\Sigma_2)$ where $J^+$ and $J^-$ denote the causal future and past, respectively.
By $\Adach$ we denote the $\Adach$-form computed from the curvature of $X$ and $\ch(\nE)$ is the Chern-character form for the curvature of $\nE$, see \cite[Sec.~4.1]{MR2273508}.
Hence $\Adach$ contains the contribution of gravitation to the relative charge and $\ch(\nE)$ that of the external field.

Moreover, $\eta(\D_{\Sigma})$ denotes the $\eta$-invariant of the Dirac operator on the Cauchy hypersurface $\Sigma$ and 
$h(\D_{\Sigma})$ the dimension of its kernel. 
The additional term $- h(\D_{\Sigma_2})$ in \eqref{eq:Q1} is caused by a different convention for the spectral projectors in \cite{Baer:2015aa} concerning the eigenvalue $0$.
There is no trangression boundary term because of the product-structure assumption near $\Sigma_1$ and $\Sigma_2$.

\subsection{Summary and examples}

We summarize the results we have obtained.

\begin{theorem}\label{hauptsatz}
Let $X$ be an even-dimensional globally hyperbolic Lorentzian spin manifold, let $\Sigma_1,\Sigma_2\subset X$ be two spacelike smooth Cauchy hypersurfaces with $\Sigma_1$ lying in the past of $\Sigma_2$.
We put $M:=J^+(\Sigma_1)\cap J^-(\Sigma_2)$.

Let $E\to X$ be a Hermitian vector bundle with compatible connection $\nE$.
We assume that the metric of $X$ and $\nE$ have product structure near $\Sigma_1$ and $\Sigma_2$.

Then the relative right-handed charge $Q_R^{\omega_{\Sigma_1},\omega_{\Sigma_2}}$ as defined in \eqref{eq:Q} for the Fock states $\omega_{\Sigma_j}$ and the Dirac operator twisted with $E$ is given by 
\begin{equation}
Q_R^{\omega_{\Sigma_1},\omega_{\Sigma_2}}
= 
-\int_M \Adach\wedge\ch(\nE) 
+\frac{h(\D_{\Sigma_1}) - h(\D_{\Sigma_2})+\eta(\D_{\Sigma_1})-\eta(\D_{\Sigma_2})}{2} \, .
\label{eq:QRFormel}
\end{equation}
\end{theorem}

Interchanging left-handed and right-handed spinors in the whole discussion we can also define the \emph{relative left-handed charge} $Q_L^{\omega_{\Sigma_1},\omega_{\Sigma_2}}$.
The projector defining the Fock representation is now given by $P=p_>(\DS^*) \oplus p_\ge(\DS)$.
Then a discussion analogous to the above yields
\begin{equation}
Q_L^{\omega_{\Sigma_1},\omega_{\Sigma_2}}
= 
\int_M \Adach\wedge\ch(\nE) 
+\frac{-h(\D_{\Sigma_1}) + h(\D_{\Sigma_2})-\eta(\D_{\Sigma_1})+\eta(\D_{\Sigma_2})}{2} \, .
\label{eq:QLFormel}
\end{equation}
The exchange of chirality is equivalent to reversing the orientation of $X$.
This explains the opposite sign for the contribution given by the integral.
The induced orientations on $\Sigma_1$ and $\Sigma_2$ will also be reversed which results in a replacement of $\D_{\Sigma_j}$ by $-\D_{\Sigma_j}$.
Hence the $\eta$-invariants get the opposite sign.
The different convention concerning the eigenvalue $0$ in the definition of the projection $P$ is responsible for the opposite signs in the $h$-terms.

If we call $Q^{\omega_{\Sigma_1},\omega_{\Sigma_2}}:=Q_R^{\omega_{\Sigma_1},\omega_{\Sigma_2}}+Q_L^{\omega_{\Sigma_1},\omega_{\Sigma_2}}$ the \emph{relative total charge} and $Q_\mathrm{chir}^{\omega_{\Sigma_1},\omega_{\Sigma_2}}:=Q_R^{\omega_{\Sigma_1},\omega_{\Sigma_2}}-Q_L^{\omega_{\Sigma_1},\omega_{\Sigma_2}}$ the \emph{relative chiral charge} then \eqref{eq:QRFormel} and \eqref{eq:QLFormel} imply
\begin{gather*}
Q^{\omega_{\Sigma_1},\omega_{\Sigma_2}} 
= 
0\, , \\
Q_\mathrm{chir}^{\omega_{\Sigma_1},\omega_{\Sigma_2}}
= 
-2\int_M \Adach\wedge\ch(\nE) 
+ h(\D_{\Sigma_1}) - h(\D_{\Sigma_2})+\eta(\D_{\Sigma_1})-\eta(\D_{\Sigma_2}) \, .
\end{gather*}
That is to say that the total charge is preserved in quantum field theory while the chiral charge is not.
The chiral anomaly depends on the space-time curvature via $\Adach$, on the external field via $\ch(\nE)$, and on spectral properties of the spatial Dirac operators via the $h$- and $\eta$-terms.

\begin{exa} The following example is well known but shows that the contributions of the $\eta$-invariant are essential to give a correct integer valued total charge.
Let $X=\mathbb{R}\times S^1$ equipped with the metric $dt^2-d\theta^2$ where $\theta$ denotes the standard coordinate on the circle $S^1=\mathbb{R}/L\mathbb{Z}$ of length $L$.
We twist with the topologically trivial complex line bundle $E$ (indeed all complex line bundles on $X$ are topologically trivial) and the connection $\nE=\partial + iA$ where the electro-magnetic potential $A$ is of the form $A=A_1(t)d\theta$.
The (real) curvature form of $\nE$ is given by $F=dA=\dot A_1 dt\wedge d\theta$.
This describes an electric field with no magnetic component.

The surface $X$ with the given metric has vanishing curvature.
But even if it had nontrivial curvature it would not enter the formula for the relative charge because $\Adach$ has nonzero contributions only in dimensions divisible by $4$.
Thus $\Adach=1$.
In two dimensions the Chern character form is simply given by $\ch(\nE) = \frac{1}{2\pi}F=\frac{\dot A_1}{2\pi}dt\wedge d\theta$.

For $t_1<t_2$ and $\Sigma_j = \{t_j\}\times S^1$ we have $M=[t_1,t_2]\times S^1$ and the integral is given by
$$
\int_M \Adach\wedge\ch(\nE) 
=
\int_{t_1}^{t_2}\int_{S^1}\frac{\dot A_1}{2\pi}\, d\theta\, dt
=
\frac{L}{2\pi}(A_1(t_2)-A_1(t_1)) \, .
$$
Now $S^1$ (and hence $X$) has two inequivalent spin structures, the \emph{trivial} and the \emph{nontrivial spin structure}.
Spinors with respect to the trivial spin structure correspond to complex-valued functions.
Spinors on $S^1$ with respect to the nontrivial spin structure correspond to functions $u$ on $\mathbb{R}$ which are antiperiodic with period $L$, i.e.\ $u(\theta+L)=-u(\theta)$.
The twisted Dirac operator on $\Sigma=\{t\}\times S^1$ takes the form $\DS=i\partial_\theta - A_1(t)$.
Its eigenvalues have multiplicity~$1$ and are given by
$$
\frac{2\pi}{L}k-A_1(t)
$$
for the trivial spin structure and
$$
\frac{2\pi}{L}\bigg(k+\frac12\bigg)-A_1(t)
$$
for the nontrivial spin structure where $k\in\mathbb{Z}$.
A little computation using Hurwitz $\zeta$-functions shows for the trivial spin structure
$$
2\cdot\frac{L}{2\pi}A_1(t) +h(\DS) +\eta(\DS)
=
2\left\lfloor\frac{L}{2\pi}A_1(t)\right\rfloor+1
$$
and hence
$$
Q_\mathrm{chir}^{\omega_{\Sigma_1},\omega_{\Sigma_2}}
=
2\left\lfloor\frac{L}{2\pi}A_1(t_1)\right\rfloor - 2\left\lfloor\frac{L}{2\pi}A_1(t_2)\right\rfloor \, .
$$
For the nontrivial spin structure we obtain
$$
Q_\mathrm{chir}^{\omega_{\Sigma_1},\omega_{\Sigma_2}}
=
2\left\lfloor\frac{L}{2\pi}A_1(t_1)-\frac12\right\rfloor - 2\left\lfloor\frac{L}{2\pi}A_1(t_2)-\frac12\right\rfloor \, .
$$
\end{exa}
\begin{exa}
Let $X=\mathbb{R}\times S^{4k-1}$ with a metric of the form $dt^2-g_t$ where $g_t$ is a one-parameter family of Riemannian metrics on $S^{4k-1}$.
The manifold $X$ has a unique spin structure.
This time we let $E$ be the trivial line bundle with trivial connection $\nE$ so that there is no external field.

We consider $M=[t_1,t_2]\times S^{4k-1}$.
It is shown in \cite[Sec.~5]{Baer:2015aa} that the family $g_t$ and the Cauchy hypersurfaces $\Sigma_j=\{t_j\}\times S^{4k-1}$ can be chosen in such a way that $h(\D_{\Sigma_1})= h(\D_{\Sigma_2})=0$ and $\int_M\Adach + \frac{\eta(\D_{\Sigma_2})-\eta(\D_{\Sigma_1})}{2}=(-1)^{k-1}{2k \choose k}$.
Hence 
$$
Q_\mathrm{chir}^{\omega_{\Sigma_1},\omega_{\Sigma_2}}
=
(-1)^k2{2k \choose k} \, .
$$
In this case the space-time metric, i.e.\ gravity, is causing the nontriviality of the chiral anomaly.
In dimension $4$ ($k=1$) this example was already discussed in \cite{gibbons1979}.
\end{exa}

\begin{exa}
Next we consider \emph{Bianchi-Type-I space-times}.
Here $X=\mathbb{R}\times \mathbb{R}^3/\Gamma$ where $\Gamma$ is a lattice in $\mathbb{R}^3$.
Denoting the coordinate of the $\mathbb{R}$-factor by $t$ and the standard coordinates of $\mathbb{R}^3$ by $x^1,x^2,x^3$ the metric of $X$ takes the form $dt^2-a_1^2(t)(dx^1)^2 - a_2^2(t)(dx^2)^2-a_3^2(t)(dx^3)^2$ where $a_i:\mathbb{R}\to\mathbb{R}$ are smooth positive functions.
Hence, for fixed $t$, the Cauchy hypersurface $\{t\}\times\mathbb{R}^3/\Gamma$ is a flat torus.

The $\Adach$-integrand can be computed to vanish identically on $X$, $\Adach\equiv0$.
The manifold $X$ has $8$ different spin structures.
The Dirac spectrum of a flat torus is symmetric about $0$ for every choice of spin spin structure so that all $\eta$-invariants vanish.
Moreover, the kernel of the Dirac operator on a flat torus is $1$-dimensional for one of the spin structures and trivial for the others.
Thus the $h(\D_{\Sigma_j})$-contributions in the formulas for the chiral anomaly cancel in all cases.
We conclude that for Bianchi-Type-I space-times the chiral anomaly vanishes,
$$
Q_R^{\omega_{\Sigma_1},\omega_{\Sigma_2}}
= 
Q_L^{\omega_{\Sigma_1},\omega_{\Sigma_2}}
=
Q_\mathrm{chir}^{\omega_{\Sigma_1},\omega_{\Sigma_2}}
=
0.
$$
\end{exa}

\begin{exa}
Finally  we consider \emph{Bianchi-Type-II space-times}.
The torus in a Bianchi-Type-I space-time is replaced by a $3$-dimensional Heisenberg manifold, i.e., $X=\mathbb{R}\times \left(\He(\mathbb{Z})\backslash\He(\mathbb{R})\right)$ where $\He(\mathbb{R})$ and $\He(\mathbb{Z})$ denotes the Heisenberg group of matrices of the form $
\begin{pmatrix}
1 & x & z \\ 
0 & 1 & y \\
0 & 0 & 1                                                                                                                                                                                                                                               \end{pmatrix}
$
with real or integral entries, respectively.
Clearly, $\He(\mathbb{R})$ is diffeomorphic to $\mathbb{R}^3$ but the quotient $\Sigma:=\He(\mathbb{Z})\backslash\He(\mathbb{R})$ is a compact $3$-manifold non-diffeomorphic to a torus.
It has four different spin structures.
For positive numbers $a$ and $b$ we obtain a left-invariant metric on $\He(\mathbb{R})$ by
$$
g_{a,b} 
= 
\bigg(\frac{b^2 y^2}{4} + a^2\bigg) dx^2 + \bigg(\frac{b^2 x^2}{4} + a^2\bigg) dy^2 + b^2 dz^2
- \frac{b^2xy}{2}dxdy + b^2ydxdz - b^2x dydz \, .
$$
By left-invariance the metric descends to a Riemannian metric on the quotient $\Sigma$.

Now we let $a,b:\mathbb{R}\to\mathbb{R}$ be smooth positive functions and we equip $X$ with the metric $dt^2-g_{a(t),b(t)}$. 
A tedious but straightforward computation yields
$$
\Adach = 
\frac{(a^2b\dot{a}^2 - a^3b\ddot{a} - a^3\dot{a}\dot{b}+ a^4\ddot{b}-b^3)(b\dot{a}-a\dot{b})}{48 \pi^2 a^5} 
dt\wedge dx\wedge dy\wedge dz \, .
$$
Integration over $M=[t_1,t_2]\times \Sigma$ yields
$$
\int_M\Adach 
=
-\left.\frac{b^4-2a^2b^2\dot{a}^2+4a^3\dot{a}b\dot{b}-2a^4\dot{b}^2}{192\pi^2a^4}\right|_{t=t_1}^{t_2}
$$
Assuming product structure near $t_1$ and $t_2$ as required in Theorem~\ref{hauptsatz}, the derivatives of $a$ and $b$ vanish at $t_1$ and at $t_2$ so that we are left with
$$
\int_M\Adach 
= 
\frac{1}{192\pi^2}\bigg(\frac{b^4(t_1)}{a^4(t_1)}-\frac{b^4(t_2)}{a^4(t_2)}\bigg) \, .
$$
The $\eta$-invariant of the Dirac operator on $\Sigma_t$ is computed in \cite[Thm.~11]{gornet-richardson} (with $A=a^2/b^2$, $r=v_1=w_2=m_v=1$ and $v_2=w_1=m_w=0$ in their notation):
$$
\eta(\D_{\Sigma_t}) = \frac{b^4(t)}{96\pi^2a^4(t)} - N(t)
$$
where $N(t)$ is an explicitly given integer depending on $a(t)/b(t)$ and the spin structure.
From the explicit formula (44) in \cite{gornet-richardson} one sees that for every choice of spin structure $N(t)\to\infty$ if  $b(t)/a(t)\to\infty$ and $N(t)\to 0$ if  $b(t)/a(t)\to 0$.
Therefore the functions $a(t)$ and $b(t)$ can be chosen such that $N(t_2)-N(t_1)$ is arbitrarily large.
Moreover, for generic choice of $a(t_j)/b(t_j)$ we have $h(\D_{\Sigma_j})=0$.
Then
$$
Q_\mathrm{chir}^{\omega_{\Sigma_1},\omega_{\Sigma_2}}
=
-2\int_M \Adach 
+ h(\D_{\Sigma_1}) - h(\D_{\Sigma_2})+\eta(\D_{\Sigma_1})-\eta(\D_{\Sigma_2}) 
=
N(t_2)-N(t_1)
$$
can take arbitrarily large positive but also arbitrarily small negative values for suitable choices of the functions $a(t)$ and $b(t)$.
\end{exa}

\section{Conclusion and outlook}

Given two Hadamard states on the field algebra for the Dirac equation, we defined a relative current which can be thought of as the expectation value (with respect to one state) of the normally ordered current operator where normal ordering has been done with respect to the other state.
In contrast to currents associated with one state, these relative currents can be defined unambigously without any need for regularization.
Integration of the relative current over a Cauchy hypersurface yields the relative charge.
In case the two states are vacuum states of observers moving along two distinguished timelike Killing fields, we expressed the relative charge in terms of an integral into which the gravitational field and the gauge field enter plus a boundary contribution essentially given by the $\eta$-invariant.
The crucial tool was a novel index theorem for hyperbolic Dirac operators which makes it possible to avoid any Wick rotation thus making the whole derivation mathematically rigorous.

Of course, the assumption on the space-time to be spatially compact is something that one would like to get rid of in order to better understand the local properties of anomalies and in order to be able to treat more models from general relativity. 
Technically, this assumption ensures that the relative charge is finite and can be expressed as an index. 
However, the relative current is locally defined and does make sense also on more general space-times. 
A local index theorem should be able to compute the local current generated by the external field. 
In case of noncompact Cauchy hypersurfaces which have compact quotients there is also the possibility to employ the concepts of $L^2$-indices, $L^2$-traces, and $L^2$-dimensions as introduced by von Neumann.

\appendix
 
\section{The canonical trace on the algebra of smoothing operators with compact support in one variable} \label{appendix:trace}

Let $X$ be a manifold with a positive volume density $\dV$. 
Suppose that $E$ is a complex vector bundle over $X$.
Then any smooth section $a$ of the bundle $E \boxtimes E^*$ will determine an operator $A: C^\infty_0(X;E) \to C^\infty(X;E)$
via
$$
 (A u)(x) = \int_M a(x,y) u(y) \;\dV(y).
$$
If the integral kernel $a$ has support contained in a set of the form $K \times X$, where $K \subset X$ is compact we say that $a$ is compactly supported in the first variable. 
The set of operators with integral kernel that is compactly supported in the first variable shall be denoted by $\Psi^{-\infty}_{0}$.
Note that this set forms an algebra and these operators map $C^\infty_0(X;E)$ to $C^\infty_0(X;E)$.
A canonical trace can be defined on this algebra by
$$
 \Tr(A):=\int_X \tr a(x,x)\; \dV(x).
$$
This trace cannot be directly interpreted as an $L^2$-trace because an operator in $\Psi^{-\infty}_{0}$ need not necessarily be $L^2$-bounded. However, given a compact exhaustion $K_n$ of $X$ and a sequence of compactly
supported smooth functions $\chi_n$ with $\chi_n(x)=1$ for all $x \in K_n$, we have
$$
 \Tr(A) = \lim_{n \to \infty} \Tr_{L^2}(A \chi_n).
$$
The operator $A \chi_n$ has smooth integral kernel $a(x,y) \chi_n(y)$ which is compactly supported in both variables.
Hence it is trace class by Mercer's theorem.
For $A,B\in\Psi^{-\infty}_{0}$ the trace property $\Tr(A B)= \Tr(B A)$ follows immediately from Fubini's theorem. 

Now let $P$ be a differential operator acting on sections of $E$.
If $A\in\Psi^{-\infty}_{0}$ then $PA,AP\in\Psi^{-\infty}_{0}$ because they have integral kernels $(P\otimes\id_{E^*})a$ and $(\id_{E}\otimes P^*)a$, respectively.
The trace property $\Tr(P A)= \Tr(A P)$ also holds even though $P\notin \Psi^{-\infty}_{0}$.
This can easily be checked using integration by parts and the fact that in local coordinates
$$
 \frac{\partial}{\partial x^i} a(x,x) = \frac{\partial}{\partial x^i} a(x,y) \vert_{y=x} + \frac{\partial}{\partial y^i} a(x,y)\vert_{y=x}.
$$

\bibliographystyle{plain}
\bibliography{lit.bib}

\end{document}